# Analysis of simultaneous 3D positioning and attitude estimation of a planar coil using inductive coupling

Antonio Moschitta, Alessio De Angelis, Marco Dionigi, Paolo Carbone
Dept. of Engineering, University of Perugia
Perugia, Italy
{antonio.moschitta,alessio.deangelis,marco.dionigi,paolo.carbone}@unipg.it

*Abstract*— In this paper, simultaneous estimation of 3D position and attitude of a single coil using a set of anchors, with known position and magnetic dipole, is analyzed. Effect of noise and geometric properties of the anchors' constellation is considered. Several parameters are analyzed and discussed, including placement of anchors in a single or in multiple orthogonal planes. It is shown that adding space and orientation diversity anchors may lead to a more robust performance when the mobile node attitude changes in time.

## I. INTRODUCTION

Magnetic fields are a well-known solution for short range positioning systems, and subject of recent research activities since they consent accurate range and position measurements. Unlike those developed using RF or ultrasound technologies, systems based on magnetic fields are robust to multipath and Line of Sight obstructions, and can be realized using low-cost electronic circuitry. Some solutions in the literature use sensors to build magnetic maps of the environment for positioning purposes, as in [1]. Better accuracy may be obtained by designing a magnetic positioning system comprising an infrastructure with artificial magnetic field sources, thus eliminating the need for extensive dataset collection [2]. Furthermore, other proposed solutions are based on DC magnetic fields or permanent magnets, as in [3]. Compared to these systems, architectures exploiting AC magnetic fields, which are typically based on coils, allow for larger operational range and may reduce power consumption if resonance is employed [4]. Furthermore, for specific applications, commercial systems using magnetic fields are available, providing high accuracy over a short range [5][6].

In this context, 3D positioning can be achieved using tri-axial coils, i.e. three coils, orthogonal to each other, in the mobile node and/or in the anchors [7][8]. However, this can limit the usability of the system, since the mobile node may not be easily manufactured or handled, unless its size is kept very small. Moreover, coil misalignments due to construction have a significant impact on positioning accuracy [9]. Note that, to mitigate such problems, a solution based on planar coils may be implemented, also using Printed Circuit Board (PCB) technology. This, in turn, may limit the operational range of the system and its accuracy, since, according to the Faraday Neumann Lenz law, the coil size affects the magnetic coupling. Consequently, systems aimed at locating single coils moving in space have been analyzed and published [10]-[13].

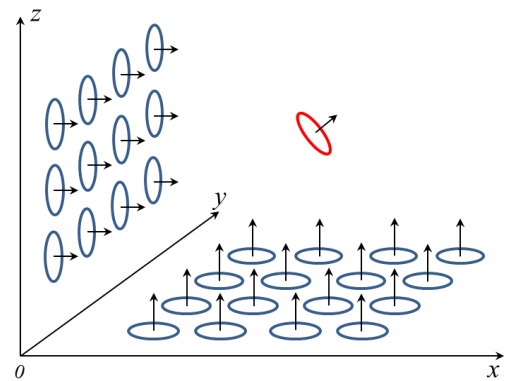

Fig. 1 – The considered system architecture. The mobile coil is represented by the red circle, while the coils acting as anchors are represented by the blue circles. For each coil, an arrow is shown, describing its orientation.

In this paper, design criteria for short range six-degrees-of-freedom (6DoF) positioning of a mobile node equipped with a simple planar coil are investigated. The considered system is assumed to rely on Received Signal Strength measurements, i.e. root mean square voltage measurement taken by the mobile node, and aims at accurately locating a node operating within less than 2 m off the beacons, as for instance in biometric applications. Various parameters are considered, that include placement of fixed anchors, Signal to Noise Ratio (SNR) at the receiver's output, and the sensitivity of the position estimation fitting to inaccurate knowledge of anchors' position and magnetic dipole moments. Note that technological parameters such as beacons' and mobile node's implementation criteria have been deeply analyzed in the literature [2]-[10]. This analysis, based on Monte Carlo simulations, aims at assessing the influence of parameters related to anchor placement, sensitivity to inaccurate knowledge of anchors parameters, and noise sensitivity. To the authors' knowledge, this issue has not been previously analyzed. In a short range system the position of the anchors can be considered as a design degree of freedom, because the environment is controlled. Consequently the proposed analysis aims at providing useful design criteria when realizing a short range positioning system based on inductive coupling of AC magnetic fields. It is shown that allocating anchors in a tridimensional pattern may result in a more robust performance with respect to placing all anchors in a single plane, which is the typical solution mentioned in the literature [10]. The rest of this paper is organized as follows. Section II describes the considered system architecture, while Section III provides simulation results. Firstly the effect on positioning accuracy of anchor placement, is studied, comparing a planar array of anchors to a 3D tri-planar array. Then the effect of inaccurate knowledge of anchors' parameters is analyzed. Finally, the effective operational area of the positioning architecture is investigated.

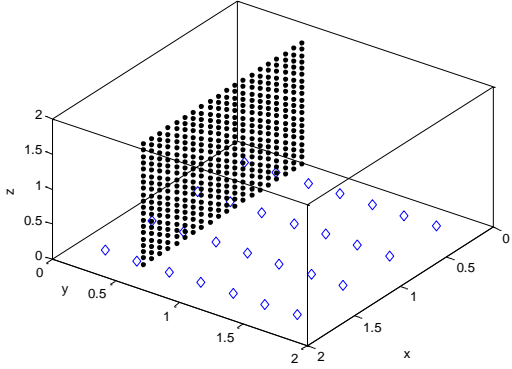

Fig. 2 – Estimation of position and attitude of a mobile coil (taking positions represented by black points, with random attitude), using a single planar array of 28 coils.

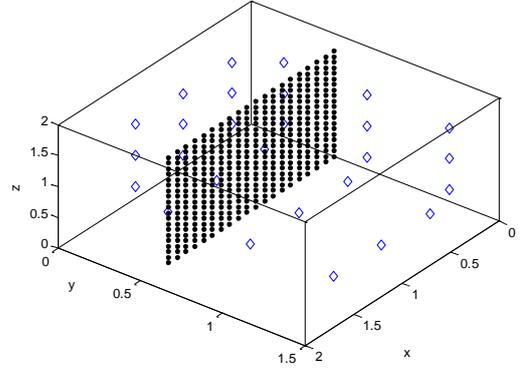

Fig. 3 – Estimation of position and attitude of a mobile coil (taking positions represented by black points, with random attitude), using 3 orthogonal planar arrays, for a total of 27 coils.

## II. SYSTEM ARCHITECTURE

The architecture of the considered system is described in Fig. 1, and consists in a set of anchors, realized by planar coils, described by their magnetic dipole moment. The mobile coil is represented as a red circle, while the beacons are represented as blue circles. Without loss of generality, throughout this paper the anchors are assumed to act as active beacons, while the mobile coil is assumed to act as a receiver. Root mean square voltage ($V_{rms}$) measurements are assumed to be collected at the mobile coil's output, using a high input impedance measuring device. Each beacon operates at a known frequency, and it is assumed that the mobile node can discriminate transmissions from different beacons. Throughout this paper, we assumed operations under steady state conditions, and that all beacon coils were stimulated by the same sinusoidal current with peak value $I_0 = I\sqrt{2}$, where $I$ is the rms value, and frequency $f_0$, given by

$$I(t) = I_0 \sin(2\pi f_0 t). \quad (1)$$

Each beacon was modeled as a coil of radius $r$ with $N_w$ windings, with center coordinates $(x_b, y_b, z_b)$. The coil orientation was described by a versor (i.e. a unit length vector) $\vec{n}_b$, modeling the direction of the coil axis. Hence, using phasor notation, the magnetic dipole moment of the $i$-th coil is given by.

$$\vec{m}_{b,i} = N_w S I \vec{n}_{b,i}, \qquad S = \pi r^2, \qquad i = 0,...,N_B - 1, \quad (2)$$

where $S$ is the coil's area, $N_B$ is the number of beacons, and $I$ is the current stimulating the coil.

By assuming known the magnetic dipole moment of each beacon, the magnetic field produced by each beacon in a given position is

$$\vec{B}_i = \frac{\mu_0}{4\pi}\left(\frac{3\vec{d}_i(\vec{m}_{b,i} \cdot \vec{d}_i)}{|\vec{d}_i|^5} - \frac{\vec{m}_{b,i}}{|\vec{d}_i|^3}\right), \quad (3)$$

where $\vec{d}_i$ is the distance vector connecting the center of the $i$-th beacon to the center of the mobile coil. Eq. (3) can be simplified as follows:

$$\vec{B}_i = \frac{\mu_0}{4\pi d_i^3}\left(3(\vec{m}_{b,i} \cdot \vec{n}_{bi,c})\vec{n}_{bi,c} - \vec{m}_{b,i}\right), \quad (4)$$

where $\vec{n}_{bi,c}$ is the unit vector associated to $\vec{d}_i$. Using (4), and assuming the mobile coil output to be connected to a high impedance measuring device, the rms voltage at the mobile coil output, induced by the $i$-th beacon, is given by

$$V_{rms,i} = 2\pi f_0 N_{w,c} S_c \vec{B}_i \cdot \vec{n}_c, \qquad i = 0,...,N_B - 1, \quad (4)$$

where $\vec{n}_c$ is a unit vector describing the mobile coil orientation. The simulation model assumes that $V_{rms,i}$ is estimated by sampling the voltage sinewave at the mobile coil's output, corrupted by an Additive White Gaussian Noise (AWGN), with standard deviation $\sigma$. Note that, in a practical scenario, the noise level can be measured when the active beacons are turned off. Thus, a set of $N_B$ noisy $V_{rms}$ measurements is collected, one for each beacon. Then, these measurements are used to evaluate the following cost function

$$F(\theta) = \sum_{i=0}^{N_F - 1}\left(\hat{V}_{rms,i} - V_{rms,i}(\theta)\right)^2, \qquad \theta = [\vec{P}, \vec{n}_e], \quad (5)$$

where the argument $\theta = [\vec{P}, \vec{n}_e]$ is composed by $\vec{P} = [x, y, z]$ and $\vec{n}_e$, that describe the mobile node position and attitude, respectively, $\hat{V}_{rms,i}$ is the measured rms voltage induced by the $i$-th beacon, and $V_{rms,i}(\theta)$ is the voltage that should be measured in absence of noise if the mobile node position and attitude were exactly $\vec{P}$ and $\vec{n}_e$. Note that, when the received signal power is very low, the resulting $V_{rms}$ measurement can be dominated by noise, and using it in (5) can reduce the accuracy of the following fitting. Consequently, a Signal to Noise Ratio (*SNR*) threshold $SNR_{th}$, expressed in dB, was defined, such that the simulator can discard noisy measurements. As a consequence, (5) evaluates a subset of $N_F$ measurements, out of the $N_B$ available ones. This procedure can also increase processing speed, because the computational complexity of the numerical fitting grows with the number of $V_{rms}$ measurements.

Table I-a

| Configuration | $P_d$ | $P_\alpha$ | $P_{d,\alpha}$ |
|---|---|---|---|
| Tri-planar | 0.984 | 1 | 0.969 |
| Mono-planar | 0.7735 | 0.9982 | 0.7398 |

Table I-b

| Configuration | $P_d$ | $P_\alpha$ | $P_{d,\alpha}$ |
|---|---|---|---|
| Tri-planar | 0.785 | 0.9995 | 0.749 |
| Mono-planar | 0.681 | 0.9955 | 0.609 |

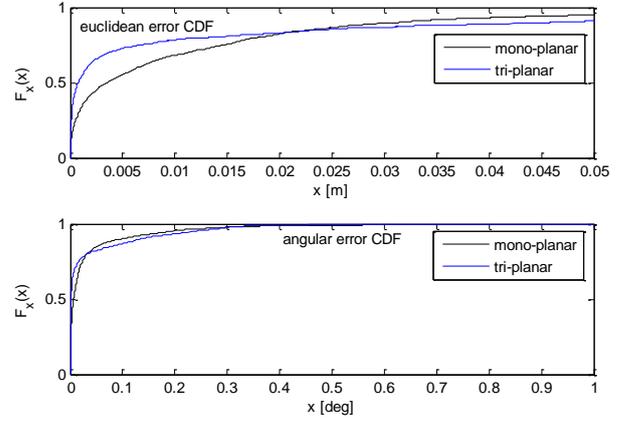

Fig. 4 – Euclidean error CDF and angular error CDF, for both mono-planar and tri-planar beacon arrays, assuming that the initial estimation of θ is affected by uncertainty.

Thus, the mobile coil position and attitude can be estimated by minimizing (5) with respect to $\vec{P}=[x,y,z]$ and $\vec{n}_e$, by means of numerical techniques. In this paper, (5) was minimized using the Nelder-Mead algorithm. Note that, being the optimization iterative, an initial estimation of the position is needed. The initialization is described in section III.B.

The simulation environment, written in Matlab, was used to test several scenarios. The results are described in the following section.

### III. ANALYSIS AND RESULTS

#### A. Test conditions and performance metrics

The simulations were initially run by assuming normalized magnetic dipole moments (i.e. with unitary magnitude), aiming at comparing the effectiveness of different configurations of anchors, and at assessing sensitivity to noise of the considered system. In a practical scenario (2) shows that such a magnetic dipole magnitude may be achieved only using a large number of windings, a large coil area, or a large stimulating current, and that a low magnitude of (2) results in low coupling with the mobile coil. Moreover, reducing the coil size may help spacing the anchors away, reducing uncertainty sources due to mutual coupling between the anchors. However, it is well known that the usage of high *Q* resonant coils can mitigate this issue when AC magnetic fields are used [4].The analysis was organized as follows: first, the effect of arranging anchors in different patterns was analyzed. Then, sensitivity to the selection of the *SNR* threshold $SNR_{th}$ was investigated. Following that, the effect of inaccuracies in the knowledge of the beacons positions and magnetic dipole moment on the fitting accuracy was analyzed (see subsection III-D). Throughout Section III, a frequency $f_0$=200 kHz was assumed, and an AWGN with σ=10μV was considered affecting the acquisition of the mobile node's signal. As a first approximation, the selected noise level is compatible with the output referred noise of an Instrumentation Amplifier, boosting the mobile coil's output.

Finally, a more realistic scenario was considered (see subsection III-E), postulating specific values for the coils radiuses, number of windings, and currents. Coupled with the selected noise, these parameters were chosen so as to describe a realistic scenario, where a mobile node operating within 2 m off the anchors is to be located.

Simulation results were compared against a set of meaningful metrics. In particular, for each considered scenario the empirical Cumulative Distribution Function (CDF) of both positioning error and attitude error were derived. The positioning error was defined as the Euclidean distance between the true mobile coil position and the estimated one, while the angular error was defined as the angle between the estimated attitude versor and the true one. Note that angular errors can take values in the [-90°, 90°] interval, because an angle of 180° does not correspond to variations in the received $V_{rms}$. As an additional performance metric, the fraction of measurements leading to a positioning error not exceeding 1 cm and to an angular error not exceeding 1° was evaluated.

#### B. Placement of anchors

In this test, two configurations of anchors were considered. With respect to a 3 axis Cartesian reference system, the first considered configuration is an array of 28 coils, lying on the *xy* plane, in a 7x4 matrix covering a square of about 1.5mx1.5m. The second considered configuration is a set of 3 planar arrays, respectively located on the *xy*, *xz*, and *yz* planes. Each planar array is a 3x3 square matrix, covering an area of about 1.2mx1.2m. In both configurations, for each array of coplanar coils, the magnetic dipole moments are orthogonal to the plane hosting the coils.

Both beacon configurations were tested against the same set of positions assumed by the mobile coil, covering a square area on a vertical plane, parallel to the *xz* plane, for a total of 400 positions. The two scenarios are shown in Fig. 2 and 3 respectively, where the beacons are shown as diamonds and the mobile coil positions are shown as black points. Note that, for each position assumed by the mobile coil, a random bearing was considered.

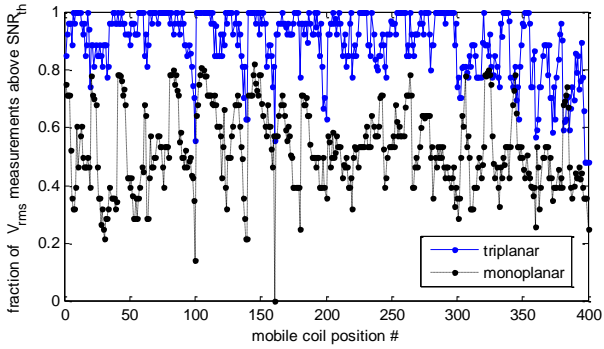

Fig. 5 - Fraction of beacons providing $V_{rms}$ measurements above a SNR threshold of 10 dB, using the mono-planar beacon array (black) and the tri-planar beacon configuration.

Table II – Euclidean and angular error statistics as a function of $SNR_{th}$.

| $SNR_{th}$ | $e_d$ mean | $e_d$ std | $e_\alpha$ mean | $e_\alpha$ std |
|---|---|---|---|---|
| 0 dB | 0.01559 | 0.044663 | 0.039248 | 0.10737 |
| 5 dB | 0.014779 | 0.040018 | 0.040072 | 0.11001 |
| 10 dB | 0.0148 | 0.0444 | 0.0373 | 0.0897 |
| 15 dB | 0.013144 | 0.040419 | 0.036089 | 0.10394 |
| 20 dB | 0.036005 | 0.16713 | 0.059954 | 0.17993 |
| 25 dB | 0.72881 | 5.6534 | 0.20672 | 0.44474 |
| 30 dB | 1,0462 | 5.1339 | 0.43236 | 0.63185 |

Under the stated conditions, two sets of simulations were run. The first one (case 'a'), aimed at assessing the robustness to AWGN of the considered configurations, was performed by using as initial condition the true position/attitude, while the second set of simulations (case 'b') was run by assuming uncertainty in the initial guess for $\theta$. To this aim, a random error, uniformly distributed in [-10 cm, 10 cm], was added to each Cartesian coordinate of the initial guess. Similarly, both the elevation the and azimuth angles associated to the attitude versor where affected by a random error, uniformly distributed in [-18°, 18°]. In particular, Table I-a and I-b show the estimated probability $P_d$ of position estimations meeting the 1 cm target for the Euclidean error $e_d$, the estimated probability $P_\alpha$ that a position measurements meets the target angular error $e_\alpha$ of 1°, and the joint probability $P_{d,\alpha}$ that a measurements meets both targets simultaneously. Table I-a was obtained under case 'a', Table I-b was obtained under case 'b'. Under both scenarios, the tri-planar configuration resulted in a better accuracy with respect to position estimation, while the two solutions are comparable when estimating the coil attitude. Fig. 4 shows the CDF of both the Euclidean and angular errors obtained under case 'b'. Using the tri-planar configuration, the Euclidean error was less than 25 cm in 99% of the simulated measurements, with a single outlier (1 case out of 4000 trials) showing $e_d$ =70cm, due to failed convergence of the fitting algorithm. When using the mono-planar configuration failed convergence occurred more frequently (4 cases out of 4000 trials) with larger positioning errors.

The reason behind the better performance of the tri-planar configurations is both in the reduced average distance between the mobile coil and the coil acting as anchors, and in the increased attitude diversity. In fact, it was observed, that, for a given position and orientation, the mobile coil may be scarcely coupled with most of the coils lying on a plane. In this case, the coil on the remaining planes are still likely to be fairly coupled to the mobile coil, because of their different orientation. Fig. 5, obtained when $SNR_{th}$=10 dB, shows the fraction of $V_{rms}$ measurements collected by using the two beacon configurations. Note that in most of the positions assumed with random attitude by the mobile coil, the tri-planar configuration leads to a larger number of beacons providing meaningful information than the mono-planar configuration.

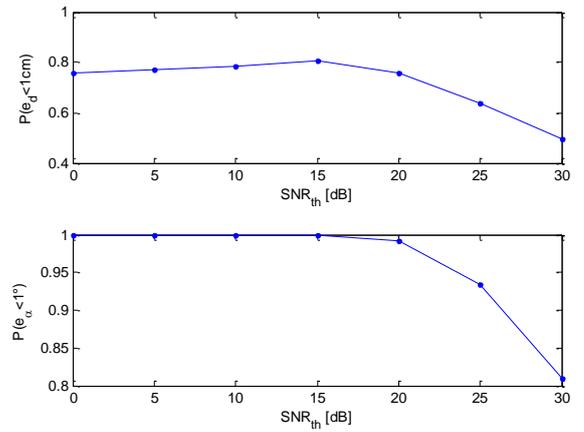

Fig. 6 – Probability that the Euclidean error $e_d$ is lower than 1cm (upper plot), and probability that the angular error $e_\alpha$ is lower than 1° (lower plot), as a function of $SNR_{th}$.

*C. SNR threshold*

Following the compared analysis in Subsection III-B, estimation of $\theta$ was further tested assuming to use the tri-planar array of anchor coils. This time, various tests were run by changing $SNR_{th}$, for the tri-planar beacon configuration. For the considered test setup, a SNR threshold of 15 dB was observed to provide optimal performance. Figs. 6, shows the probability that Euclidean and angular error are less than 1cm and 1° respectively, and in Table II, that shows mean value and standard deviation of both Euclidean and angular error. Note that the existence of an optimal value $SNR_{th}$ was expected. In fact, for low values of $SNR_{th}$ many $V_{rms}$ measurements dominated by noise make their way into the numerical fitting algorithm, increasing the noise effect on the estimation of $\theta$. Conversely, when a high value of $SNR_{th}$ is selected, fewer and fewer $V_{rms}$ measurements are deemed eligible for fitting, possibly discarding measurements carrying useful information for the estimation of $\theta$.

*D. Sensitivity to inaccurate knowledge of anchors*

The sensitivity to anchors' parameters was investigated using Monte Carlo simulation. At each iteration, when simulating the measurement of $V_{rms}$, the software introduced a random uniformly distributed perturbation in the magnetic dipole moments associated to each beacon, while the numerical minimization of (5) used the nominal magnetic dipole moment values when evaluating $V_{rms,i}(\theta)$.

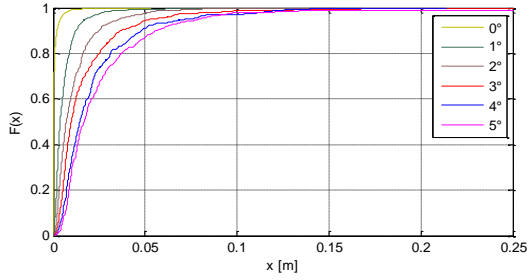

Fig. 7 – Euclidean error CDF obtained by assuming a random uniformly distributed error in the attitude of the beacons' magnetic dipole moments, upper bounded by 1°, 2°, 3°, 4°, and 5° respectively. The error-free case is shown for comparison purposes.

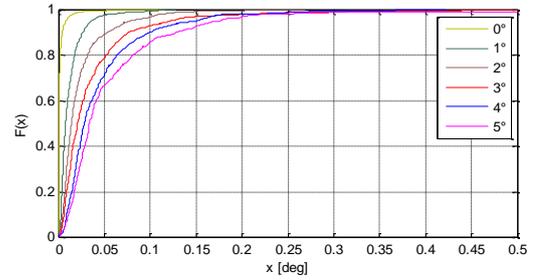

Fig. 8 – Angular error CDF obtained by assuming a random uniformly distributed error in the attitude of the beacons' magnetic dipole moments, upper bounded by 1°, 2°, 3°, 4°, and 5° respectively. The error-free case is shown for comparison purposes.

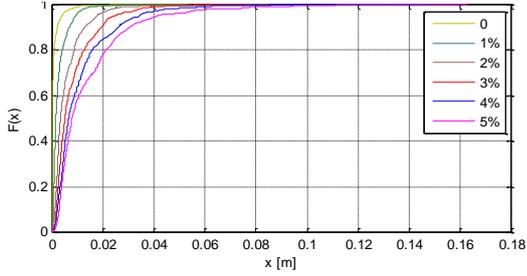

Fig. 9 – Euclidean error CDF obtained by assuming a random uniformly distributed error in the magnitude of the beacons' magnetic dipole moments, upper bounded by 1%, 2%, 3%, 4%, and 5% respectively. The error-free case is shown for comparison purposes.

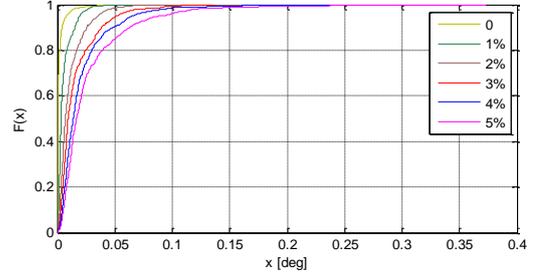

Fig. 10 – Angular error CDF obtained by assuming a random uniformly distributed error in the magnitude of the beacons' magnetic dipole moments, upper bounded by 1%, 2%, 3%, 4%, and 5% respectively. The error-free case is shown for comparison.

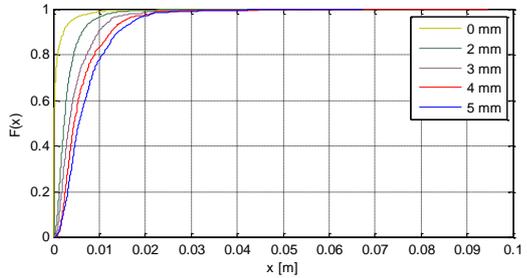

Fig. 11 – Euclidean error CDF obtained by assuming a random uniformly distributed error in the position of the beacons, upper bounded by 1 mm, 2 mm, 3 mm, 4 mm, and 5 mm respectively. The error-free case is shown for comparison purposes.

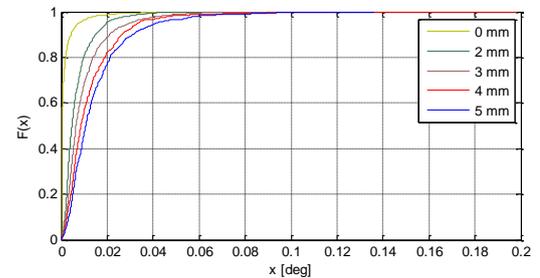

Fig. 12 – Angular error CDF obtained by assuming a random uniformly distributed error in the position of the beacons, upper bounded by 1 mm, 2 mm, 3 mm, 4 mm, and 5 mm respectively. The error-free case is shown for comparison purposes.

This procedure was used to simulate the effect of inaccurate knowledge of point of application, direction and magnitude of beacons' magnetic dipole moments. Errors in the point of application can model errors in beacons' placement, while errors in magnetic dipole moments bearing can model coils fabrication tolerances, and magnitude errors can describe uncertainty in current feeding the coils. The Monte Carlo analysis, covering the scenario in Fig. 3 under case 'a' conditions, led to the results summarized in Figs. 7-12, that show Euclidean and angular error CDF under various conditions. In particular, Fig. 7 shows that the positioning performance is most sensitive to errors in the magnetic dipole direction, since for a maximum angular error of 5° $P_d$ drops from 0.98 to 0.27. Fig. 9 shows that the positioning performance is fairly sensitive to errors in the magnetic dipole magnitude, since a for maximum magnitude error of 5% $P_d$ drops from 0.98 to 0.58. Finally, Fig. 11 shows that the positioning performance is fairly tolerant to beacon placement errors. In fact, when a maximum placement error of 5 mm on the beacons' coordinates is considered, $P_d$ drops from 0.98 to 0.77. Similar conclusions can be drawn for the angular error, comparing Figs. 8, 10, and 12. However, the system angular accuracy is seemingly more robust to inaccurate knowledge of beacons, since the observed angular error was always lower than 0.5°.

### E. Analysis of a realistic scenario

Finally, beacon and mobile nodes were modeled in greater detail, assuming beacons to be 20 winding coils with a radius of 3 cm, fed by a 2 A rms current (this value may be achieved using high-Q resonant coils). The mobile node was modeled by a 10 winding coil with a radius of 1 cm, followed by an amplifier with gain *G*. Accurate knowledge of the anchors' parameters was assumed. In order to investigate the stability of the numerical fitting, the Nelder-Mead algorithm was fed with a fixed number $N$ of measurements, obtained by taking the $N$ largest values among the 27 $V_{rms}$ measurements collected by the mobile node. Simulations were initially run, under the scenario depicted in Fig. 3, assuming $G$=20 (i.e. 26 dB), and $N$=7. A performance comparable to the error free case in Figs. 7-12 was observed, with an Euclidean mean error of 1.8 cm and an angular mean error of 0.02°. However, by repeating the simulation with larger values of $N$, failed convergence was often observed, with large positioning errors. To gain further insight, the positions corresponding to Euclidean errors larger than 1 cm or to angular errors larger than 1° were shown as red circlets in Fig. 13, obtained for $N$=10. Fig. 13 shows that large errors mostly occur when the mobile node is far from the majority of the anchors, leading only a few anchors of the tri-planar grid to provide useful information for fitting purposes. Hence simulations were repeated by assuming a smaller operational area, shown in Fig. 14, for various values of $G$. This time, the fitting was fed with up to 12 $V_{rms}$ measurements, chosen among those exceeding $SNR_{th}$. As shown in Tab. III, in all considered cases a mean error of less than 1 cm was observed.

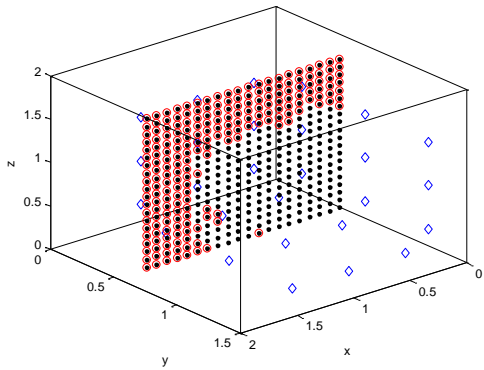

Fig. 13 – Mobile node positions leading to positioning outliers (red circlets), obtained by assuming realistic coils when simulating the scenario of Fig. 3.

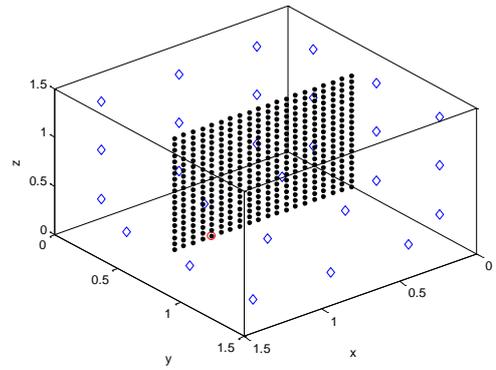

Fig. 14 – Positioning outliers (red circlets), obtained by assuming realistic coils when simulating the scenario of Fig. 3, when the mobile node takes positions closer to the area covered by the beacons.

Table III – Euclidean and angular error statistics as a function of $G$, under the conditions of Fig. 14

| $G$ | $e_d$ mean | $e_d$ std | $e_\alpha$ mean | $e_\alpha$ std |
|---|---|---|---|---|
| 1 | 0.0059 | 0.03 | 0.09 | 0.93 |
| 5 | 0.015 | 0.0024 | 0.0024 | 0.03 |
| 10 | 0.012 | 0.022 | 0.0018 | 0.035 |
| 15 | 0.012 | 0.022 | 0.0013 | 0.027 |
| 20 | 0.012 | 0.024 | 0.0014 | 0.029 |

## IV. CONCLUSIONS

This paper is focused on criteria to design and optimize a positioning system based on inductive coupling, capable of estimating both the position and the attitude of a single coil. The placement of anchors was discussed, showing that placing all anchors on different orthogonal planes may lead to increased accuracy and robustness. The selection of collected measurement on a SNR basis was investigated as well. Finally, sensitivity to inaccuracies in realizing the anchor coils was investigated, keeping into account placement, orientation, and current amplitude errors. The operation of a realistic system was also simulated.

ACKNOWLEDGEMENT

This research activity was funded through grant PRIN 2015C37B25 by the Italian Ministry of Instruction, University and Research (MIUR), whose support the authors gratefully acknowledge.